\documentclass[preprint,showpacs,preprintnumbers,amsmath,amssymb]{revtex4}

\usepackage{graphicx}
\usepackage{dcolumn}
\usepackage{bm}
\usepackage{hyperref}
\hypersetup{colorlinks=false}

\begin{document}


\title{Bound State Calculations of the Three-Dimensional Yakubovsky Equations with the inclusion of Three-Body Forces}

\author{M.~R. Hadizadeh}
\email{hadizade@khayam.ut.ac.ir}
\author{S. Bayegan}%
 \email{bayegan@khayam.ut.ac.ir}
\affiliation{%
Department of Physics, University of Tehran, P.O.Box 14395-547,
Tehran, Iran
}%

\date{\today}

\begin{abstract}

The four-body Yakubovsky equations in a
Three-Dimensional approach with the inclusion of the three-body
forces is proposed. The four-body bound state with two- and
three-body interactions is formulated in Three-Dimensional
approach for identical particles as function of vector Jacobi
momenta, specifically the magnitudes of the momenta and the angles
between them. The modified three dimensional Yakubovsky integral
equations is successfully solved with the scalar two-meson
exchange three-body force where the Malfliet-Tjon-type two-body
force is implemented. The three-body force effects on the energy
eigenvalue and the four-body wave function, as well as accuracy of
our numerical calculations are presented.The four-body Yakubovsky equations in a
Three-Dimensional approach with the inclusion of the three-body
forces is proposed. The four-body bound state with two- and
three-body interactions is formulated in Three-Dimensional
approach for identical particles as function of vector Jacobi
momenta, specifically the magnitudes of the momenta and the angles
between them. The modified three dimensional Yakubovsky integral
equations is successfully solved with the scalar two-meson
exchange three-body force where the Malfliet-Tjon-type two-body
force is implemented. The three-body force effects on the energy
eigenvalue and the four-body wave function, as well as accuracy of
our numerical calculations are presented.
\end{abstract}

\maketitle

\section{Introduction}

The topic of three-body forces (3BFs) is as old as nuclear physics
\cite{Primakoff-PR55} and, based on meson exchanges, various
processes have been proposed in the past (for a review see
\cite{Robilotta-FBS35}). Among them, the Fujita-Miyazawa force
\cite{Fujita-PTP17} with an intermediate $\Delta$ generated by the
exchange of two pions is most obvious and is implemented in all
modern 3BF models. However the nature of these 3BFs is still not
completely understood. In recent years there has been new progress
in understanding the form of nuclear forces, because of the
application of chiral perturbation theory ($\chi$PT)
\cite{Meissner-NPA684}-\cite{Epelbaum-PLB639}. From this
developments one can expect a more systematic understanding of the
form of two-body (2B) and 3B forces. However, $\chi$PT implies
\textit{a priori} unknown constants, the low-energy constants,
which have to be determined from experimental data. The bound
states of few-nucleons seem to be an ideal laboratory to determine
3BF parameters, as the binding energies are sensitive to the 3B
interaction and they are expected to be governed by the low-energy
regime of nuclear physics \cite{Epelbaum-PPNP57,Bedaque-NPA676}.
Therefore, the understanding of nuclear few-body bound states is
an important contribution to the understanding of the 3BF. To this
aim and for their numerical investigations one requires reliable
methods leading to the solutions of the non-relativistic
Schr\"{o}dinger equation.

In the past several solution methods have been developed and
applied to the four-body bound state problem by using realistic
nuclear potentials, the CRCGV \cite{Hiyama-PRL85}, the SV
\cite{Usukura-PRB59}, the HH \cite{Viviani-PRC71}, the GFMC
\cite{Viringa-PRC62}, the NCSM \cite{Navratil-PRC62}, EIHH
\cite{Barnea-PRC67} and the Faddeev-Yakubovsky (FY)
\cite{Schellingerhout-PRC46}-\cite{Epelbaum-PRC70}. These
calculational schemes are mostly based on a partial wave (PW)
decomposition. Stochastic and Monte Carlo methods, however, are
performed directly using position vectors in configuration space.
One of the most viable approaches appears to be the FY method. The
calculations based on FY are performed after a PW expansion with
phenomenological potentials in configuration space
\cite{Schellingerhout-PRC46,Lazauskas-PRC7}, and in momentum space
\cite{Kamada-NPA548}-\cite{Nogga-PRC65} and recently with chiral
potentials in momentum space
\cite{Epelbaum-PRL86}-\cite{Epelbaum-PRC70}.

The FY scheme based on a PW decomposition, which includes spin and
isospin degrees of freedom, after truncation leads to two coupled
sets of a finite number of coupled equations in three variables
for the amplitudes. In PW decomposition the number of channels
that must be included grows very rapidly in this case, and a
further complication is arisen where there are now six spatial
dimensions rather than the three required for three-body
calculations. So in a PW decomposition one needs a tremendous
number of partial waves to find converged results. In view of this
very large number of interfering terms it appears natural to give
up such an expansion and work directly with vector variables.

On this basis recently we have extended the Three-Dimensional (3D)
approach, which greatly simplifies the two- and three-body
scattering and bound state calculations without using PW
decomposition \cite{Elster-FBS24}-\cite{Lin-PRC76}, to the
four-body bound state \cite{Hadizadeh-WS,Hadizadeh-FBS40}. We have
formulated the Yakubovsky equations with only 2BFs as function of
vector Jacobi momenta, specifically the magnitudes of the momenta
and the angles between them. We have obtained two coupled
three-dimensional integral equations in six variables for the
amplitudes which greatly simplifies the calculations without using
PW decomposition. The obtained three-dimensional integral
equations have been solved successfully for simple NN force
models. In this paper we follow the same approach and consider the
3BFs in four-body bound state problem. As a simplification we
neglect spin and isospin degrees of freedom and study the
four-boson bound state problem.

So the purpose of this work is to demonstrate that one can solve
the Yakubovsky equations for four-body bound state without using
PW decomposition and in the first attempt we have done it by using
very simple 2B and 3B model interactions.

In our formulation we work directly with vector variables in the
Yakubovsky scheme in momentum space. Here the dependence on
momentum vectors shows that our 3D representation in comparison to
traditional PW representation avoids the very involved angular
momentum algebra occurring for the permutations and especially for
the 3BFs and the full solution can be reached exactly and simply
whereas the PW representation of the amplitudes leads to rather
complicated expressions \cite{Nogga-PHD}.

We believe that this work is another step forward in the
development of 3D approach for studying the few-body systems and
it is the first attempt towards the solution of the 4N bound state
problem with the inclusion of 3NFs without performing the PW
decomposition.

This paper is organized as follows. In section \ref{section: YEs}
we briefly represent the coupled Yakubovsky equations for
four-body bound state with two- and three-body interactions. In
section \ref{section: 3BF} we evaluate the matrix elements of
3BFs. In section \ref{section: coordinate systems} we discuss our
choice for independent variables for the unknown amplitudes in the
equations and in their kernels. Section \ref{section: numerical
techniques} describes details of our algorithm for solving coupled
Yakubovsky three-dimensional integral equations. In section
\ref{section: numerical results} we present our results for three-
and four-body binding energies with and without model 3BFs and we
provide the test of our calculation. Finally we summarize in
section \ref{section: summary} and provide an outlook.

\section{Momentum Space Representation of Yakubovsky Equations with 3BFs}\label{section: YEs}

\begin{figure}
  \includegraphics{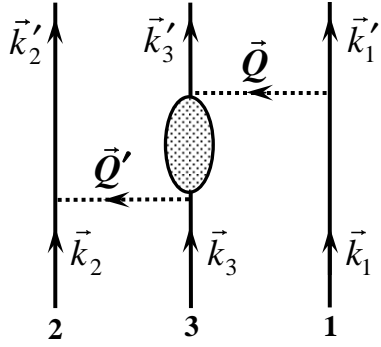}
\caption{Diagrammatic representation of the part $W_{123}^{(3)}$
of a two-meson exchange 3BF. Here particle 3 is single out by the
meson-nucleon amplitude described by the blob.}
\label{fig:3BF}       
\end{figure}

The bound state of the four-body (4B) system, in the presence of
3BFs, is described by two coupled Yakubovsky equations
\cite{Nogga-PRC65}:
\begin{eqnarray}
\label{Eq.YCs} |\psi_{1}\rangle &=& G_{0}tP
[(1+P_{34})|\psi_{1}\rangle+|\psi_{2}\rangle] +(1+G_{0}t) G_{0}
W_{123}^{(3)} |\Psi\rangle \nonumber \\* |\psi_{2}\rangle &=&
G_{0}t\tilde{P}[(1+P_{34})|\psi_{1}\rangle+|\psi_{2}\rangle]
\end{eqnarray}
where the Yakubovsky components $|\psi_{1}\rangle$ and
$|\psi_{2}\rangle$ belong to $"3+1" (123,4;12)$ and $"2+2"
(12,34;12)$ partitions of the four particles respectively. Here
the free four-body propagator is given by $G_{0}=(E-H_{0})^{-1}$,
and $H_{0}$ stands for the free Hamiltonian. The operator $t$ is
the two-body transition matrix determined by a two-body
Lippman-Schwinger equation. $P$, $\tilde{P}$ and $P_{34}$ are
permutation operators. $P=P_{12}P_{23}+P_{13}P_{23}$ permutes the
particles in three-body subsystem (123) and
$\tilde{P}=P_{13}P_{24}$ interchanges the two two-body subclusters
(12) and (34). The quantity $W_{123}^{(3)}$, as shown in
Fig.~\ref{fig:3BF}, defines a part of the 3BF in the cluster
$(123)$, which is symmetric under the exchange of particles $1$
and $2$ and which can be related by an interchange of the three
particles to two other parts $W_{123}^{(1)}$ and $W_{123}^{(2)}$
that sum up to the total 3BF of particles 1, 2 and 3:
$W_{123}=W_{123}^{(1)}+W_{123}^{(2)}+W_{123}^{(3)}$. The total 4B
wave function $|\Psi\rangle$ is given as:

\begin{equation}
|\Psi\rangle=(1+P+P_{34}P+\tilde{P})[(1+P_{34})|\psi_{1}\rangle+|\psi_{2}\rangle]
\label{Eq.WF}
\end{equation}

\begin{figure}
  \includegraphics{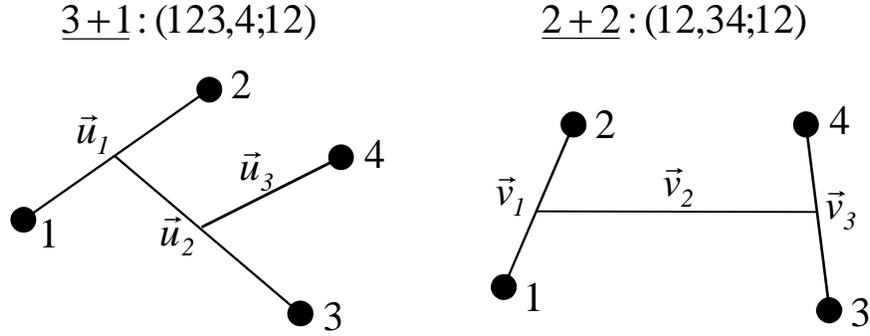}
\caption{Definition of the $3+1$ and $2+2$ type of Jacobi
coordinates.}
\label{fig:basis states}       
\end{figure}

The symmetry property of $|\psi_{1}\rangle$ under exchange of
particles $1$ and $2$, and $|\psi_{2}\rangle$ under separate
exchanges of particles $1,2$ and $3,4$ guarantee that
$|\Psi\rangle$ is totally symmetric. It can easily be verified
that the inclusion of the 3BF component $W_{123}^{(3)}$ into the
definition of the first Yakubovsky component $|\psi_{1}\rangle$
does not change its symmetry property.

In this paper we follow the notation introduced in Ref.
\cite{Hadizadeh-FBS40} and work in a 3D momentum space basis.
According to the two types of chains $(123,4;12)$ and $(12,34;12)$
there are two type of basis states, Fig.~\ref{fig:basis states},
which are suitable to represent the two Yakubovsky components
$|\psi_{1}\rangle$ and $|\psi_{2}\rangle$ in the coupled equations
(\ref{Eq.YCs}). The representation of coupled equations
(\ref{Eq.YCs}) in these basis sates will be exactly the same as
displayed in Ref. \cite{Hadizadeh-FBS40} except that an extra term
with $W_{123}^{(3)}$ occurs in the first component. This is
\begin{eqnarray}
  \langle \vec{u}_{1}\,\vec{u}_{2}\,\vec{u}_{3}| (1+G_{0}t_{12}) G_{0} W_{123}^{(3)} |\Psi\rangle
  &&  \nonumber
\\* && \hspace{-35mm} =\int
D^{3}\tilde{u} '\, \langle \vec{u}_{1}\,\vec{u}_{2}\,\vec{u}_{3}|
(1+G_{0}t_{12}) G_{0}
|\vec{\tilde{u}}\,'_{1}\,\vec{\tilde{u}}\,'_{2}\,\vec{\tilde{u}}\,'_{3}\rangle
 \nonumber
\\* && \hspace{-35mm} \times
\langle\vec{\tilde{u}}\,'_{1}\,\vec{\tilde{u}}\,'_{2}\,\vec{\tilde{u}}\,'_{3}|W_{123}^{(3)}
|\Psi\rangle
   \label{Eq.13}
\end{eqnarray}
where $D^{3}u \equiv d^{3}u_{1}\, d^{3}u_{2}\, d^{3}u_{3}$. The
first matrix element can be handled as described in Ref.
\cite{Hadizadeh-FBS40}. The second matrix element involves the
3BF, which has been worked out in Ref. \cite{Liu-FBS33} in a 3D
momentum space basis for three-body system. After evaluating the
first matrix element in Eq.~(\ref{Eq.13}), the coupled three
dimensional Yakubovsky integral equations can be rewrite
explicitly as:

\begin{eqnarray}
\langle \vec{u}_{1}\,\vec{u}_{2}\,\vec{u}_{3}|\psi_{1}\rangle &=&
\frac{1}{{E-\frac{u_{1}^{2}}{m}
-\frac{3u_{2}^{2}}{4m}-\frac{2u_{3}^{2}}{3m}}} \nonumber
\\* && \hspace{-30 mm} \times \Biggl [\, \int d^{3}u_{2}'
\,\, \langle\vec{u}_{1}|t_{s}(\epsilon) |\frac{1}{2}\vec{u}_{2}
+\vec{u}\,'_{2} \rangle   \nonumber \\*  && \hspace{-20mm} \times
\Biggl \{\,\, \langle\vec{u}_{2}+\frac{1}{2} \vec{u}\,'_{2} \,\,
\vec{u}\,'_{2}\,\,\vec{u}_{3}|\psi_{1}\rangle  \nonumber
\\* \quad && \hspace{-14mm} +\langle\vec{u}_{2}+\frac{1}{2}
\vec{u}\,'_{2} \,\, \frac{1}{3}\vec{u}\,'_{2}+
\frac{8}{9}\vec{u}_{3} \,\,
\vec{u}\,'_{2}-\frac{1}{3}\vec{u}_{3}|\psi_{1}\rangle
 \nonumber \\*  && \hspace{-14mm} +
 \langle\vec{u}_{2}+\frac{1}{2} \vec{u}\,'_{2}\,\, -\vec{u}\,'_{2}-\frac{2}{3}\vec{u}_{3} \,\,
\frac{1}{2}\vec{u}\,'_{2}-\frac{2}{3}\vec{u}_{3}
|\psi_{2}\rangle\,\, \Biggr\} \nonumber \\*  && \hspace{-27mm} +
 \Biggl \{\langle\vec{u}\,_{1}\,\vec{u}\,_{2}\,\vec{u}\,_{3}|W_{123}^{(3)}
|\Psi\rangle    \nonumber \\* && \hspace{-24mm}+ \frac{1}{2} \int
d^{3}\tilde{u}_{1}'\, \frac{\langle \vec{u}_{1}|
t_{s}(\epsilon)|\vec{\tilde{u}}\,'_{1}\rangle}{E-\frac{\tilde{u}_{1}'^{2}}{m}
-\frac{3u_{2}^{2}}{4m}-\frac{2u_{3}^{2}}{3m}}
\langle\vec{\tilde{u}}\,'_{1}\,\vec{u}\,_{2}\,\vec{u}\,_{3}|W_{123}^{(3)}
|\Psi\rangle \Biggr\} \, \, \Biggr]
 \nonumber \\* \nonumber \\* \nonumber \\*
\langle \vec{v}_{1}\,\vec{v}_{2}\,\vec{v}_{3}|\psi_{2}\rangle &=&
\frac{\frac{1}{2} \int d^{3}v_{3}' \,
\langle\vec{v}_{1}|t_{s}(\epsilon^{*})| \vec{v}\,'_{3}\rangle\,
}{E-\frac{v_{1}^{2}}{m} -\frac{v_{2}^{2}}{2m}-\frac{v_{3}^{2}}{m}}
\nonumber \\* && \hspace{-25 mm} \times  \Biggl\{\, 2\,
\langle\vec{v}_{3}\,\,
 \frac{2}{3}\vec{v}_{2}+\frac{2}{3}\vec{v}\,'_{3} \,\, \frac{1}{2}\vec{v}_{2}-\vec{v}\,'_{3} |\psi_{1}\rangle
  + \langle\vec{v}_{3}\,\,-\vec{v}_{2}\,\,
\vec{v}\,'_{3}|\psi_{2}\rangle \, \Biggr\}  \nonumber \\*
\label{Eq.14}
 \end{eqnarray}
where the $t_{s}(\epsilon)$ and $t_{s}(\epsilon^{*})$ are
symmetrized two-body $t$-matrices with the two-body subsystem
energies $\epsilon =
E-\frac{3u_{2}^{2}}{4m}-\frac{2u_{3}^{2}}{3m}$ and
$\epsilon^{*}=E-\frac{v_{2}^{2}}{2m}-\frac{v_{3}^{2}}{m}$. The
matrix elements of the 3BF term,
$\langle\vec{u}\,_{1}\,\vec{u}\,_{2}\,\vec{u}\,_{3}|W_{123}^{(3)}
|\Psi\rangle$, are evaluated in the next section.

\section{The Evaluation of 3BF Matrix Elements in a 3D Approach}\label{section: 3BF}
Each part of a 3BF with two scalar meson exchanges and a constant
meson-nucleon amplitude, which is shown in Fig.~\ref{fig:3BF}, can
be written in the following form
\begin{eqnarray}
W_{123}^{(3)} \propto\frac{F(Q\,^{2})}{Q\,^{2}+m^{2}_{s}} \:
\frac{F(Q'\,^{2})}{Q'\,^{2}+m^{2}_{s}} \label{Eq.15}
\end{eqnarray}
with a cutoff function
\begin{equation}
F(Q\,^{2}) =
\Biggl(\frac{\Lambda^{2}-m_s^{2}}{\Lambda^{2}+Q\,^{2}}\Biggr)^{2}
\label{Eq.16}
\end{equation}
and momentum transfers $\vec{Q}$ and $\vec{Q}'$
\begin{eqnarray}
\vec{Q}&=&\vec{k}_{1}-\vec{k}'_{1}   \nonumber \\ &\equiv& \Biggl
\{(+\vec{u}_{1}-\frac{1}{2}\vec{u}_{2})-(+\vec{u}\,'_{1}-\frac{1}{2}\vec{u}\,'_{2})\Biggr\}_{(123,4;12)} \nonumber \\
&\equiv& \{\vec{u}_{2}-\vec{u}\,'_{2}\}_{(231,4;23)} \nonumber \\ \nonumber \\
\vec{Q}'&=&\vec{k}_{2}'-\vec{k}_{2}  \nonumber \\ &\equiv& \Biggl\{
(-\vec{u}\,'_{1}-\frac{1}{2}\vec{u}\,'_{2})-(-\vec{u}_{1}-\frac{1}{2}\vec{u}_{2})\Biggr\}_{(123,4;12)}
\nonumber \\ &\equiv& \{\vec{u}\,'_{2}-\vec{u}\,_{2}\}_{(312,4;31)}
\label{Eq.17}
\end{eqnarray}
where the multiple indices for each curly bracket denote the
two-body followed by the $3+1$ fragmentation.

For the evaluation of  Eq.~(\ref{Eq.14}) matrix elements of the
form
$\langle\vec{u}\,_{1}\,\vec{u}\,_{2}\,\vec{u}\,_{3}|W_{123}^{(3)}
|\Psi\rangle$ need to be calculated. From Fig.~\ref{fig:3BF} we
see that $W_{123}^{(3)}$ can be considered as a sequence of meson
exchanges in the subsystem (23), where it is called for
convenience subsystem 1, and subsystem (31), is called 2. Since
the structure of the 3BF we consider is specified by two momentum
transfers of consecutive meson exchanges, it is convenient to
insert a complete set of states of the type $2$ between
$W_{123}^{(3)}$ and $|\Psi\rangle$ and another complete set of
states of type $1$ between the two meson exchanges. Then the
matrix element of $W_{123}^{(3)}$ is rewritten as
\begin{eqnarray}
&&
_{3}\langle\vec{u}\,_{1}\,\vec{u}\,_{2}\,\vec{u}\,_{3}|W_{123}^{(3)}
|\Psi\rangle    \nonumber \\ && \hspace{0mm} = \int_{1} D^{3} u'
\,\, _{3}\langle\vec{u}\,_{1}\,\vec{u}\,_{2}\,\vec{u}\,_{3}|
 \vec{u}\,'_{1}\,\vec{u}\,'_{2}\,\vec{u}\,'_{3} \rangle_{1}
\nonumber \\ && \hspace{0mm} \times \int_{1} D^{3} u'' \,\,
_{1}\langle
\vec{u}\,'_{1}\,\vec{u}\,'_{2}\,\vec{u}\,'_{3}|\frac{F(Q\,^{2})}{Q\,^{2}+m^{2}_{s}}|
\vec{u}\,''_{1}\,\vec{u}\,''_{2}\,\vec{u}\,''_{3} \rangle_{1}
\nonumber \\ && \hspace{0mm} \times
 \int_{2} D^{3} u''' \,\, _{1}\langle\vec{u}\,''_{1}\,\vec{u}\,''_{2}\,\vec{u}\,''_{3}|
 \vec{u}\,'''_{1}\,\vec{u}\,'''_{2}\,\vec{u}\,'''_{3} \rangle_{2}
 \nonumber \\ && \hspace{0mm} \times \int_{2} D^{3} u'''' \,\,
_{2}\langle
\vec{u}\,'''_{1}\,\vec{u}\,'''_{2}\,\vec{u}\,'''_{3}|\frac{F(Q'\,^{2})}{Q'\,^{2}+m^{2}_{s}}|
\vec{u}\,''''_{1}\,\vec{u}\,''''_{2}\,\vec{u}\,''''_{3}\rangle_{2}
\,  \nonumber \\ && \hspace{0mm} \times \, _{2}\langle
\vec{u}\,''''_{1}\,\vec{u}\,''''_{2}\,\vec{u}\,''''_{3}|\Psi\rangle
 \label{Eq.18}
\end{eqnarray}
Here the subscripts $1, 2, 3$ of the bra and ket vectors and in
integrals stand for the different types of three-body coordinate
systems of $(3+1)$-type fragmentation $(ijk,4;ij)$. Both
meson-exchange propagators in the 3BF term only depend on the
momentum transfer in a two-body subsystem, as indicated in
Eq.~(\ref{Eq.17}), i.e.
\begin{eqnarray}
&& _{1}\langle
\vec{u}\,'_{1}\,\vec{u}\,'_{2}\,\vec{u}\,'_{3}|\frac{F(Q\,^{2})}{Q\,^{2}+m^{2}_{s}}|
\vec{u}\,''_{1}\,\vec{u}\,''_{2}\,\vec{u}\,''_{3} \rangle_{1}
\nonumber \\ && \hspace{+1mm} = \Biggl\{
\delta^{3}(\vec{u}\,'_{1}-\vec{u}\,''_{1})
\delta^{3}(\vec{u}\,'_{3}-\vec{u}\,''_{3})
\frac{F((\vec{u}\,'_{2}-\vec{u}\,''_{2})^{2})}{(\vec{u}\,'_{2}-\vec{u}\,''_{2})^{2}+m^{2}_{s}} \Biggr\}_{1}
\nonumber \\
\nonumber \\
&& _{2}\langle
\vec{u}\,'''_{1}\,\vec{u}\,'''_{2}\,\vec{u}\,'''_{3}|\frac{F(Q'\,^{2})}{Q'\,^{2}+m^{2}_{s}}|
\vec{u}\,''''_{1}\,\vec{u}\,''''_{2}\,\vec{u}\,''''_{3}\rangle_{2}
\nonumber \\ && \hspace{+1mm} =  \Biggl\{
\delta^{3}(\vec{u}\,'''_{1}-\vec{u}\,''''_{1})
\delta^{3}(\vec{u}\,'''_{3}-\vec{u}\,''''_{3})
\frac{F((\vec{u}\,''''_{2}-\vec{u}\,'''_{2})^{2})}{(\vec{u}\,''''_{2}-\vec{u}\,'''_{2})^{2}+m^{2}_{s}}
\Biggr\}_{2} \nonumber \\ \label{Eq.19}
\end{eqnarray}
Using Eq.~(\ref{Eq.19}), one can rewrite Eq.~(\ref{Eq.18}) as:

\begin{eqnarray}
&&
_{3}\langle\vec{u}\,_{1}\,\vec{u}\,_{2}\,\vec{u}\,_{3}|W_{123}^{(3)}
|\Psi\rangle  \nonumber \\ && \hspace{0mm}= \int_{1} D^{3} u' \,\,
_{3}\langle\vec{u}\,_{1}\,\vec{u}\,_{2}\,\vec{u}\,_{3}|
 \vec{u}\,'_{1}\,\vec{u}\,'_{2}\,\vec{u}\,'_{3} \rangle_{1}
\nonumber \\ && \hspace{0mm}\ \times \int_{1} d^{3} u''_{2}\, \,\,
\Biggl[\frac{F((\vec{u}\,'_{2}-\vec{u}\,''_{2})^{2})}{(\vec{u}\,'_{2}-\vec{u}\,''_{2})^{2}+m^{2}_{s}}\Biggr]_{1}
\nonumber \\ && \hspace{0mm}\ \times
 \int _{2} D^{3} u''' \,\,_{1}\langle\vec{u}\,'_{1}\,\vec{u}\,''_{2}\,\vec{u}\,'_{3}|
 \vec{u}\,'''_{1}\,\vec{u}\,'''_{2}\,\vec{u}\,'''_{3} \rangle_{2}
 \nonumber \\ && \hspace{0mm}\ \times \int_{2} d^{3} u''''_{2} \,\,
\Biggl[\frac{F((\vec{u}\,''''_{2}-\vec{u}\,'''_{2})^{2})}{(\vec{u}\,''''_{2}-\vec{u}\,'''_{2})^{2}+m^{2}_{s}}
\Biggr]_{2} \nonumber \\ && \hspace{0mm}\ \times \,\,_{2} \langle
\vec{u}\,'''_{1}\,\vec{u}\,''''_{2}\,\vec{u}\,'''_{3}|\Psi\rangle
 \label{Eq.20}
\end{eqnarray}

We would like to point out that in our vector based method the
calculation of the transformations from one three-body subsystem to
another, i.e. $_{3}\langle\,|
 \, \rangle_{1}$ and $_{1}\langle\,|
 \, \rangle_{2}$, are efficiently five-dimensional interpolations, whereas in
calculation of the coordinate transformations via a PW
decomposition, there is a complicated angular momentum recoupling
algebra involved.

Also we would like to mention that we do not follow the explicit
evaluation of the coordinate transformations in Eq.~(\ref{Eq.20})
leading to expressions with meson propagators which contain linear
combinations of three or four momentum vectors. Thus direct
integrations for evaluating the matrix element of the 3BF would
involve magnitudes of momentum vectors and angles between all of
them, which can be very complicated and involved. We therefore
follow the method proposed in Ref. \cite{Liu-FBS33} and do not
carry out the coordinate transformation analytically, we evaluate
the integration of Eq.~(\ref{Eq.20}) in separate steps where in
each step we only integrate over one vector variable at a time.
 Thus we define an auxiliary function
\begin{eqnarray}
F_{2}(\vec{u}\,'''_{1},\vec{u}\,'''_{2},\vec{u}\,'''_{3}) &=&
\int_{2} d^{3} u''''_{2} \,\,
\Biggl[\frac{F((\vec{u}\,''''_{2}-\vec{u}\,'''_{2})^{2})}{(\vec{u}\,''''_{2}-\vec{u}\,'''_{2})^{2}+m^{2}_{s}}
\Biggr]_{2} \nonumber \\ &\times& \,\,_{2} \langle
\vec{u}\,'''_{1}\,\vec{u}\,''''_{2}\,\vec{u}\,'''_{3}|\Psi\rangle
\label{Eq.22}
\end{eqnarray}
the integration of the meson exchange between particles 2 and 3 in
Eq.~(\ref{Eq.22}) is carried out completely in the coordinate
system of type $2$. Once
$F_{2}(\vec{u}\,'''_{1},\vec{u}\,'''_{2},\vec{u}\,'''_{3})$ is
obtained, it needs to be expressed in terms of momenta in a
coordinate system of type 1 in order to carry out the integration
over the remaining meson exchange.  This transformation, labeled
$F_{21}(\vec{u}\,'_{1},\vec{u}\,''_{2},\vec{u}\,'_{3})$ is
explicitly given as
\begin{eqnarray}
&& F_{21}(\vec{u}\,'_{1},\vec{u}\,''_{2},\vec{u}\,'_{3})  \nonumber
\\ && = \int _{2} D^{3} u'''
\,\,_{1}\langle\vec{u}\,'_{1}\,\vec{u}\,''_{2}\,\vec{u}\,'_{3}|
 \vec{u}\,'''_{1}\,\vec{u}\,'''_{2}\,\vec{u}\,'''_{3} \rangle_{2}
\, F_{2}(\vec{u}\,'''_{1},\vec{u}\,'''_{2},\vec{u}\,'''_{3})  \nonumber \\
&&=
F_{2}(-\frac{1}{2}\vec{u}\,'_{1}-\frac{3}{4}\vec{u}\,''_{2},\vec{u}\,'_{1}-\frac{1}{2}\vec{u}\,''_{2},\vec{u}\,'_{3})
\label{Eq.23}
\end{eqnarray}
Here we used that
$F_{2}(\vec{u}\,'''_{1},\vec{u}\,'''_{2},\vec{u}\,'''_{3})$ is a
scalar function due to the total wave function
$\Psi(\vec{u}_{1}\,\vec{u}_{2}\,\vec{u}_{3})$ being a scalar in
the ground state. In our vector based method, this transformation
is effectively a five dimensional interpolation on $F_{2}$ in
Eq.~(\ref{Eq.22}), which can be handled by the cubic Hermitian
splines of Ref. \cite{Huber-FBS22}. The integration over the
second  meson exchange between particle $3$ and $1$ in the
coordinate system of type $1$ is now given by
\begin{eqnarray}
 F_{1}(\vec{u}\,'_{1},\vec{u}\,'_{2},\vec{u}\,'_{3}) && \nonumber
\\ && \hspace{-25mm} = \int_{1} d^{3} u''_{2}\, \,\,
\{\frac{F((\vec{u}\,'_{2}-\vec{u}\,''_{2})^{2})}{(\vec{u}\,'_{2}-\vec{u}\,''_{2})^{2}+m^{2}_{s}}\}_{1}
\,\,F_{21}(\vec{u}\,'_{1},\vec{u}\,''_{2},\vec{u}\,'_{3})
\hspace{+5mm} \label{Eq.24}
\end{eqnarray}
The matrix element
$_{3}\langle\vec{u}\,_{1}\,\vec{u}\,_{2}\,\vec{u}\,_{3}|W_{123}^{(3)}
|\Psi\rangle$ is finally obtained by integrating
$F_{1}(\vec{u}\,'_{1},\vec{u}\,'_{2},\vec{u}\,'_{3})$ over
$\vec{u}\,'_{1}, \vec{u}\,'_{2}$ and $\vec{u}\,'_{3}$, i.e.
carrying out the final coordinate transformation from the system
of type $1$ back to the one of type $3$,
\begin{eqnarray}
&&
_{3}\langle\vec{u}\,_{1}\,\vec{u}\,_{2}\,\vec{u}\,_{3}|W_{123}^{(3)}
|\Psi\rangle \nonumber \\ &&= \int_{1} D^{3} u' \,\,
_{3}\langle\vec{u}\,_{1}\,\vec{u}\,_{2}\,\vec{u}\,_{3}|
 \vec{u}\,'_{1}\,\vec{u}\,'_{2}\,\vec{u}\,'_{3} \rangle_{1} \,\, F_{1}(\vec{u}\,'_{1},\vec{u}\,'_{2},\vec{u}\,'_{3})
 \nonumber \\ &&= F_{1}(-\frac{1}{2}\vec{u}_{1}-\frac{3}{4}\vec{u}_{2},\vec{u}_{1}-\frac{1}{2}\vec{u}_{2},\vec{u}_{3}) \label{Eq.25}
\end{eqnarray}

\section{Choosing The Coordinate Systems}\label{section: coordinate systems}
In order to solve the coupled three dimensional Yakubovsky integral
equations, Eq.~(\ref{Eq.14}), directly without employing PW
projection, we have to define suitable coordinate systems. The
Yakubovsky components are given as a function of Jacobi momenta
vectors and as a solution of integral equations. Since we ignore
spin and isospin dependencies, the both Yakubovsky components are
scalars and thus only depend on the magnitudes of Jacobi momenta and
the angles between them. The first important step for an explicit
calculation is the selection of independent variables. As indicated
in Ref. \cite{Liu-FBS33} one needs six variables to uniquely specify
the geometry of the three vectors. The coupled three dimensional
Yakubovsky integral equations, Eq.~(\ref{Eq.14}), with only 2BFs was
solved successfully in Ref. \cite{Hadizadeh-FBS40}. For the
evaluation of the 3BF term in the first Yakubovsky component in
Eq.~(\ref{Eq.14}),
$_{3}\langle\vec{u}\,_{1}\,\vec{u}\,_{2}\,\vec{u}\,_{3}|W_{123}^{(3)}
|\Psi\rangle$, we start with calculating first
$F_{2}(\vec{u}\,'''_{1},\vec{u}\,'''_{2},\vec{u}\,'''_{3})$,
Eq.~(\ref{Eq.22}), and realize that for this integration we can
choose $\vec{u}\,'''_{3}$ parallel to the $z$-axis and
$\vec{u}\,'''_{2}$ in the $x-z$ plane. This leads to the
simplification of the azimuthal angles. The explicit expression is
\begin{eqnarray}
&&
F_{2}(u'''_{1},u'''_{2},u'''_{3},x'''_{1},x'''_{2},x_{u'''_{1}u'''_{2}}^{u'''_{3}})
 \nonumber \\ &=& \int^{\infty}_{0} du_{2}''''
u_{2}''''\,^{2}\int^{+1}_{-1}dx_{2}''''\int^{2\pi}_{0}d\phi_{2}''''
\,\,\, \Gamma(u''''_{2},u'''_{2},y_{2'''2''''}) \nonumber \\
&& \hspace{0mm}  \times
\Psi(u'''_{1},u''''_{2},u'''_{3},x'''_{1},x''''_{2},x_{u'''_{1}u''''_{2}}^{u'''_{3}})
\label{Eq.26}
\end{eqnarray}
with
\begin{eqnarray*}
 u'''_{1}&=&|\vec{u}\,'''_{1}|
\nonumber\\* u'''_{2}&=&|\vec{u}\,'''_{2}| \nonumber\\*
u'''_{3}&=&|\vec{u}\,'''_{3}| \nonumber\\*
u''''_{2}&=&|\vec{u}\,''''_{2}| \nonumber\\*
x'''_{1}&=&\hat{u}'''_{3}.\hat{u}'''_{1}\equiv
\cos(\vartheta'''_{1})\nonumber\\*
x'''_{2}&=&\hat{u}'''_{3}.\hat{u}'''_{2} \equiv
\cos(\vartheta'''_{2})\nonumber\\*
x''''_{2}&=&\hat{u}'''_{3}.\hat{u}''''_{2} \equiv
\cos(\vartheta''''_{2}) \label{Eq.27}
\end{eqnarray*}

\begin{eqnarray}
 y_{1'''2'''}&=&\hat{u}'''_{1}.\hat{u}'''_{2} \nonumber\\* &\equiv&
x'''_{1}x'''_{2}+\sqrt{1-x_{1}'''^{2}}\sqrt{1-x_{2}'''^{2}}\cos(\varphi'''_{1})\nonumber\\*
  y_{1'''2''''}&=&\hat{u}'''_{1}.\hat{u}''''_{2} \nonumber\\* &\equiv&
x'''_{1}x''''_{2}+\sqrt{1-x_{1}'''^{2}}\sqrt{1-x_{2}''''^{2}}\cos(\varphi'''_{1}-\varphi''''_{2})\nonumber\\*
 y_{2'''2''''}&=&\hat{u}'''_{2}.\hat{u}''''_{2} \nonumber\\* &\equiv&
x'''_{2}x''''_{2}+\sqrt{1-x_{2}'''^{2}}\sqrt{1-x_{2}''''^{2}}\cos(\varphi''''_{2})\nonumber\\*
x_{u'''_{1}u'''_{2}}^{u'''_{3}}&=&\frac{y_{1'''2'''}-x'''_{1}x'''_{2}}{\sqrt{1-x_{1}'''^{2}}\sqrt{1-x_{2}'''^{2}}}
\nonumber\\*
x_{u'''_{1}u''''_{2}}^{u'''_{3}}&=&\frac{y_{1'''2''''}-x'''_{1}x''''_{2}}{\sqrt{1-x_{1}'''^{2}}\sqrt{1-x_{2}''''^{2}}}
\nonumber\\*
\Pi^{2}&=&u_{2}''''^{2}+u_{2}'''^{2}+2u''''_{2}u'''_{2}y_{2'''2''''}
\nonumber\\* && \hspace{-16mm}
\Gamma(u''''_{2},u'''_{2},y_{2'''2''''})=
\frac{F(u''''_{2},u'''_{2},y_{2'''2''''})}{\Pi^{2}+m^{2}_{s}}\
\label{Eq.27}
\end{eqnarray}

Similarly for the integration over the second meson exchange, i.e.,
the evaluation of
$F_{1}(\vec{u}\,'_{1},\vec{u}\,'_{2},\vec{u}\,'_{3})$ of
Eq.~(\ref{Eq.24}), we can choose $\vec{u}\,'_{3}$ parallel to the
$z$-axis and $\vec{u}\,'_{2}$ in the $x-z$ plane. This leads to the
explicit expression which is functionally the same as
Eq.~(\ref{Eq.26}):
\begin{eqnarray}
&&
F_{1}(u'_{1},u'_{2},u'_{3},x'_{1},x'_{2},x_{u'_{1}u'_{2}}^{u'_{3}})
\nonumber \\ && = \int^{\infty}_{0} du_{2}''
u_{2}''\,^{2}\int^{+1}_{-1}dx_{2}''\int^{2\pi}_{0}d\phi_{2}'' \,\,
\Gamma(u'_{2},u''_{2},y_{2'2''}) \nonumber \\ && \times
F_{21}(u'_{1},u''_{2},u'_{3},x'_{1},x''_{2},x_{u'_{1}u''_{2}}^{u'_{3}})
\label{Eq.28}
\end{eqnarray}
with the same variables as Eq.~(\ref{Eq.27}) with $u'_{1}, u'_{2},
u'_{3}, u''_{2}, x'_{1},$ $x'_{2}, x''_{2}, \varphi'_{1},
\varphi''_{2}$ instead of $u'''_{1}, u'''_{2}, u'''_{3}, u''''_{2},
x'''_{1}, x'''_{2}, x''''_{2}, \varphi'''_{1},$ $\varphi''''_{2}$.
The evaluation of
$F_{21}(\vec{u}\,'_{1},\vec{u}\,''_{2},\vec{u}\,'_{3})$,
Eq.~(\ref{Eq.23}), is not an integration but rather a five
dimensional interpolation and explicitly is given by
\begin{eqnarray}
F_{21}(u'_{1},u''_{2},u'_{3},x'_{1},x''_{2},x_{u'_{1}u''_{2}}^{u'_{3}})
&& \nonumber \\  && \hspace{-30mm} =
F_{2}(\Pi_{1},\Pi_{2},u'_{3},x_{\Pi_{1}u'_{3}},x_{\Pi_{2}u'_{3}},x_{\Pi_{1}\Pi_{2}}^{u'_{3}})
\label{Eq.29}
\end{eqnarray}
with
\begin{eqnarray}
 \Pi_{1}&=&|-\frac{1}{2}\vec{u}\,'_{1}-\frac{3}{4}
\vec{u}\,''_{2}|=\frac{1}{2}\sqrt{u_{1}'^{2}+\frac{9}{4}u_{2}''^{2}+3u'_{1}u''_{2}y_{1'2''}}
\nonumber
\\*  \Pi_{2}&=&|\vec{u}\,'_{1}- \frac{1}{2}\vec{u}\,''_{2}|=\sqrt{u_{1}'^{2}+\frac{1}{4}u_{2}''^{2}-u_{1}'u_{2}''y_{1'2''}} \nonumber \\*
x_{\Pi_{1}u'_{3}}&=&(\widehat{-\frac{1}{2}\vec{u}\,'_{1}-\frac{3}{4}\vec{u}\,''_{2}}
) . \hat{u}\,'_{3} =\frac{1}{\Pi_{1}}(-\frac{1}{2}u'_{1}
x'_{1}-\frac{3}{4}u''_{2}x''_{2}) \nonumber
\\*
x_{\Pi_{2}u'_{3}}&=&(\widehat{\vec{u}\,'_{1}-\frac{1}{2}\vec{u}\,''_{2}}
) .\hat{u}\,'_{3} =\frac{1}{\Pi_{2}}(u'_{1}
x'_{1}-\frac{1}{2}u''_{2}x''_{2})
 \nonumber
\\*
x_{\Pi_{1}\Pi_{2}}&=&(\widehat{-\frac{1}{2}\vec{u}\,'_{1}-\frac{3}{4}\vec{u}\,''_{2}}
). (\widehat{\vec{u}\,'_{1}-\frac{1}{2}\vec{u}\,''_{2}} ) \nonumber
\\* &=&
\frac{1}{\Pi_{1}\Pi_{2}}(-\frac{1}{2}
u_{1}'^{2}+\frac{3}{8}u_{2}''^{2}-\frac{1}{2}u'_{1}u''_{2}y_{1'2''})
\nonumber
\\*
x_{\Pi_{1}\Pi_{2}}^{u'_{3}}
&=&\frac{x_{\Pi_{1}\Pi_{2}}-x_{\Pi_{1}u'_{3}}x_{\Pi_{2}u'_{3}}}
{\sqrt{1- x_{\Pi_{1}u'_{3}}^{2}}\sqrt{1-x_{\Pi_{2}u'_{3}}^{2}}}
\label{Eq.30}
\end{eqnarray}
Finally, the matrix element
$_{3}\langle\vec{u}\,_{1}\,\vec{u}\,_{2}\,\vec{u}\,_{3}|W_{123}^{(3)}
|\Psi\rangle$ is explicitly obtained by a five dimensional
interpolation as
\begin{eqnarray}
_{3}\langle\vec{u}\,_{1}\,\vec{u}\,_{2}\,\vec{u}\,_{3}|W_{123}^{(3)}
|\Psi\rangle &=&
F_{1}(\pi_{6},\pi_{7},u_{3},x_{\pi_{6}u_{3}},x_{\pi_{7}u_{3}},x_{\pi_{6}\pi_{7}}^{u_{3}})
\nonumber
\\* \label{Eq.31} \end{eqnarray} with
\begin{eqnarray}
 \pi_{6}&=&|-\frac{1}{2}\vec{u}_{1}-\frac{3}{4}
\vec{u}_{2}|=\frac{1}{2}\sqrt{u_{1}^{2}+\frac{9}{4}u_{2}^{2}+3u_{1}u_{2}y_{12}}
\nonumber
\\*  \pi_{7}&=&|\vec{u}_{1}- \frac{1}{2}\vec{u}_{2}|=\sqrt{u_{1}^{2}+\frac{1}{4}u_{2}^{2}-u_{1}u_{2}y_{12}} \nonumber \\*
x_{\pi_{6}u_{3}}&=&(\widehat{-\frac{1}{2}\vec{u}_{1}-\frac{3}{4}\vec{u}_{2}}
) .\hat{u}_{3} =\frac{1}{\pi_{6}}(-\frac{1}{2}u_{1}
x_{1}-\frac{3}{4}u_{2}x_{2}) \nonumber
\\*
x_{\pi_{7}u_{3}}&=&(\widehat{\vec{u}_{1}-\frac{1}{2}\vec{u}_{2}} )
.\hat{u}_{3} =\frac{1}{\pi_{7}}(u_{1} x_{1}-\frac{1}{2}u_{2}x_{2})
 \nonumber
\\*
x_{\pi_{6}\pi_{7}}&=&(\widehat{-\frac{1}{2}\vec{u}_{1}-\frac{3}{4}\vec{u}_{2}}
). (\widehat{\vec{u}_{1}-\frac{1}{2}\vec{u}_{2}} ) \nonumber
\\* &=&
\frac{1}{\pi_{6}\pi_{7}}(-\frac{1}{2}
u_{1}^{2}+\frac{3}{8}u_{2}^{2}-\frac{1}{2}u_{1}u_{2}y_{12})
\nonumber
\\*
x_{\pi_{6}\pi_{7}}^{u_{3}}
&=&\frac{x_{\pi_{6}\pi_{7}}-x_{\pi_{6}u_{3}}x_{\pi_{7}u_{3}}}
{\sqrt{1- x_{\pi_{6}u_{3}}^{2}}\sqrt{1-x_{\pi_{7}u_{3}}^{2}}}
\label{Eq.32}
\end{eqnarray}
The last term of first Yakubovsky component in Eq.~(\ref{Eq.14})
requires an additional integration of the matrix element
$\langle\vec{u}_{1}\,\vec{u}_{2}\,\vec{u}_{3}|W_{123}^{(3)}
|\Psi\rangle$ and the fully off-shell two-body $t$-matrix. Again,
with choosing $\vec{u}\,_{3}$ parallel to the $z$-axis we only have
four vectors to consider, $\vec{u}_{1}, \vec{u}_{2}, \vec{u}_{3}$
and $\vec{u}\,'_{1}$, thus the integration is of a similar type as
the one of the first three terms of first Yakubovsky component in
Eq.~(\ref{Eq.14}),

\begin{eqnarray}
&& \frac{1}{2} \int d^{3}\tilde{u}_{1}'\, \frac{\langle \vec{u}_{1}|
t_{s}(\epsilon)|\vec{\tilde{u}}\,'_{1}\rangle}{E-\frac{\tilde{u}_{1}'^{2}}{m}
-\frac{3u_{2}^{2}}{4m}-\frac{2u_{3}^{2}}{3m}}
\langle\vec{\tilde{u}}\,'_{1}\,\vec{u}\,_{2}\,\vec{u}\,_{3}|W_{123}^{(3)}
|\Psi\rangle  \nonumber
\\*  && = \frac{1}{2} \int^{\infty}_{0} d\tilde{u}_{1}'
\tilde{u}_{1}'\,^{2}\int^{+1}_{-1}d\tilde{x}_{1}'\int^{2\pi}_{0}d\tilde{\phi}_{1}'
\frac{ t_{s}(u_{1},\tilde{u}'_{1},y_{\tilde{1}'1};\epsilon)
}{E-\frac{\tilde{u}_{1}'^{2}}{m}
-\frac{3u_{2}^{2}}{4m}-\frac{2u_{3}^{2}}{3m}} \nonumber
\\*  && \times
\langle\vec{\tilde{u}}\,'_{1}\,\vec{u}\,_{2}\,\vec{u}\,_{3}|W_{123}^{(3)}
|\Psi\rangle    \label{Eq.33}
 \end{eqnarray}
with
\begin{eqnarray}
 \tilde{u}'_{1}&=&|\vec{\tilde{u}}\,'_{1}| \nonumber\\*
\tilde{x}'_{1}&=&\hat{u}_{3}.\hat{\tilde{u}}'_{1}\equiv
\cos(\tilde{\vartheta}'_{1})\nonumber\\*
y_{\tilde{1}'1}&=&\hat{\tilde{u}}'_{1}.\hat{u}_{1}\equiv
\tilde{x}'_{1}x_{1}+\sqrt{1-\tilde{x}_{1}'^{2}}\sqrt{1-x_{1}^{2}}\cos(\tilde{\varphi}'_{1}-\varphi_{1})\nonumber\\*
 y_{\tilde{1}'2}&=&\hat{\tilde{u}}'_{1}.\hat{u}_{2}\equiv
\tilde{x}'_{1}x_{2}+\sqrt{1-\tilde{x}_{1}'^{2}}\sqrt{1-x_{2}^{2}}\cos(\tilde{\varphi}'_{1})
\label{Eq.34}
\end{eqnarray}
These considerations lead to the explicit representation for the
Yakubovsky components $|\psi_{1}\rangle $ and $|\psi_{2}\rangle$:

\begin{eqnarray}
\psi_{1}(u_{1}\,u_{2}\,u_{3}\,x_{1}\,x_{2}\, x_{12}^{3} )  &=&
\frac{1}{{E-\frac{u_{1}^{2}}{m}
-\frac{3u_{2}^{2}}{4m}-\frac{2u_{3}^{2}}{3m}}} \nonumber
\\* && \hspace{-42mm}\times \Biggl [\, \int_{0}^{\infty}
du'_{2} \, u'^{2}_{2} \int_{-1}^{+1} dx'_{2} \int_{0}^{2\pi}
d\varphi'_{2} \,\, t_{s}(u_{1},\tilde{\pi},\tilde{x} ;\epsilon) \,
\nonumber \\* && \hspace{-34mm} \times \Biggl \{\,\,
\psi_{1}(\pi_{1}\,\, u_{2}'\,\, {u}_{3}\,\,
x_{12}\,\,x_{13}\,\,x_{\pi_{1}u_{2}'}^{u_{3}}) \nonumber
\\* \quad && \hspace{-27mm} +
\psi_{1}(\pi_{1}\,\, \pi_{2}\,\, \pi_{3}\,\,
x_{22}\,\,x_{23}\,\,x_{\pi_{1}\pi_{2}}^{\pi_{3}})
 \nonumber \\*  && \hspace{-27mm} +
\psi_{2}(\pi_{1}\,\, \pi_{4}\,\, \pi_{5}\,\,
x_{32}\,\,x_{33}\,\,x_{\pi_{1}\pi_{4}}^{\pi_{5}})\,\, \Biggr \}
\nonumber
\\*  && \hspace{-35mm} +
F_{1}(\pi_{6},\pi_{7},u_{3},x_{\pi_{6}u_{3}},x_{\pi_{7}u_{3}},x_{\pi_{6}\pi_{7}}^{u_{3}})
\nonumber
\\*  && \hspace{-35mm} +\frac{1}{2} \int^{\infty}_{0} d\tilde{u}_{1}'
\tilde{u}_{1}'\,^{2}\int^{+1}_{-1}d\tilde{x}_{1}'\int^{2\pi}_{0}d\tilde{\phi}_{1}'
  \frac{
t_{s}(u_{1},\tilde{u}'_{1},y_{\tilde{1}'1};\epsilon)
}{E-\frac{\tilde{u}_{1}'^{2}}{m}
-\frac{3u_{2}^{2}}{4m}-\frac{2u_{3}^{2}}{3m}} \nonumber \\* &&
\hspace{-30mm} \times
 F_{1}(\pi_{6}',\pi_{7}',u_{3},x_{\pi_{6}'u_{3}},x_{\pi_{7}'u_{3}},x_{\pi_{6}'\pi_{7}'}^{u_{3}})
\, \, \Biggr ]
 \nonumber \\* \nonumber \\*  \nonumber \\*
\psi_{2}(v_{1}\,v_{2}\,v_{3}\,X_{1}\,X_{2}\, X_{12}^{3} ) &=&
\frac{\frac{1}{2}}{E-\frac{v_{1}^{2}}{m}
-\frac{v_{2}^{2}}{2m}-\frac{v_{3}^{2}}{m}}  \nonumber \\* &&
\hspace{-42mm}\times \int_{0}^{\infty} dv'_{3} \, v'^{2}_{3}
\int_{-1}^{+1} dX_{3}' \int_{0}^{2\pi} d\phi'_{3} \,\,
t_{s}(v_{1},v_{3}',Y_{13^{'}};\epsilon^{*})
 \nonumber \\* && \hspace{-37mm} \times \Biggl \{\, 2\,
\psi_{1}(v_{3}\,\,\Sigma_{1} \,\,\Sigma_{2} \,\,
X_{12}\,\,X_{13}\,\,X_{v_{3}\Sigma_{1}}^{\Sigma_{2}})
   \nonumber \\* && \hspace{-32mm}  + \psi_{2}(v_{3}\,\,v_{2}
\,\,v_{3}' \,\, X_{22}\,\,X_{23}\,\,X_{v_{3}v_{2}}^{v_{3}'}) \,
\Biggr \} \label{Eq.35}
 \end{eqnarray}
The coupled equations, Eq.~(\ref{Eq.35}), is the starting point
for numerical calculations and the details will be described in
the next section. The 3D representation of total wave function
$|\Psi \rangle$ which directly appears in Eqs.~(\ref{Eq.26}) and
(\ref{Eq.35}) is represented in Ref. \cite{Hadizadeh-FBS40}, where
we have presented it as function of vector Jacobi momenta.

In a standard PW representation Eq.~(\ref{Eq.14}) is replaced by
two coupled sets of a finite number of coupled integral equations
\cite{Glockle-NPA560}, where the evaluation of two-body
$t-$matrices and permutation operators $P, \tilde{P}$ and $P_{34}$
as well as coordinate transformations due to considering angular
momentum quantum numbers instead of angle variables leads to more
complicated expressions in comparison to our 3D representation.

\section{Numerical Techniques}\label{section: numerical techniques}
In this section we describe some details of the numerical
algorithm for solving the coupled Yakubovsky three dimensional
integral equations, and more details can be found in Ref.
\cite{Hadizadeh-FBS40}. The Yakubovsky components are given as a
function of Jacobi momenta vectors and as a solution of coupled
three dimensional integral equations, Eq.~(\ref{Eq.35}). The both
Yakubovsky components are scalars and thus only depend on the
magnitudes of Jacobi momenta and the angles between them. The
dependence on the continuous momentum and angle variables should
be replaced in the numerical treatment by a dependence on certain
discrete values. For this purpose we use the Gaussian quadrature
grid points. The coupled Yakubovsky equations represent a set of
three dimensional homogenous integral equations, which after
discreatization turns into a huge matrix eigenvalue equation. The
huge matrix eigenvalue equation requires an iterative solution
method. We use a Lanczos-like scheme that is proved to be very
efficient for nuclear few-body problems \cite{Stadler-PRC44}. The
momentum variables have to cover the interval $[0,\infty]$. In
practice we limit the intervals to suitable cut-offs and their
values are chosen large enough to achieve cut-off independence.
The functional behavior of the kernel of eigenvalue equation is
determined by the two-body $t-$matrices. We also solve the
Lippman-Schwinger equation for the fully-off-shell two-body
$t-$matrices directly as function of the Jacobi vector variables
\cite{Elster-FBS24}. Since the coupled integral equations require
a very large number of interpolations, we use the cubic Hermitian
splines of Ref. \cite{Huber-FBS22} for its accuracy and high
computational speed. It should be mentioned that by adding the
additional grid points, $0$ to all momentum and $\pm1$ to all
angle grid points, we avoid the extrapolation outside the Gaussian
grids.

\section{Numerical Results}\label{section: numerical results}

\subsection{Three- and Four-Body Binding Energies}

In our calculations for 2BF we employ the spin-averaged
Malfliet-Tjon V potential \cite{Malfliet-NPA127}. This force is a
superposition of a short-ranged repulsive and long-ranged
attractive Yukawa interactions. We use the same parameters as
given in Ref. \cite{Hadizadeh-FBS40} where the nucleon mass is
defined by $\frac{\hbar^{2}}{m}=41.470 \,{\mathrm{MeV \,
fm^{2}}}$. With this interaction we solve the Lippman-Schwinger
equation for the fully-off-shell two-body $t-$matrices directly as
function of the Jacobi vector variables as described in Ref.
\cite{Elster-FBS24}. The so obtained $t-$matrices are then
symmetrized to get $t_{s}(u_{1},\tilde{\pi},\tilde{x} ;\epsilon)$
and $t_{s}(v_{1},v_{3}',Y_{13^{'}};\epsilon^{*})$.

\begin{table}
\caption{\label{table1}Three-body binding energies with and without
three-body forces in MeV. The numbers in parenthesis are binding
energies calculated in Ref. $^{33}$ for three-body bound state with
a modified version of Malfliet-Tjon by a cutoff function of dipole
type. Also the number in bracket is calculated in FY scheme in PW
representation, Ref. \cite{Kamada-NPA548}.}
\begin{tabular}{cc}
\hline
Potential  & Three-body Binding Energy  \\
\hline
MT-V  &  -7.74  [-7.73]  \\
MT-V+MT3-I     &  -8.92  \\
MT-V+MT3-II    &  -8.70  \\
MT2-II  &  -7.69  (-7.70)  \\
MT2-II+MT3-I     &  -8.87 (-8.87)  \\
MT2-II+MT3-II    &  -8.64 (-8.65)  \\
\hline
\end{tabular}
\end{table}

\begin{table}
\caption{\label{table2}Four-body binding energies with and without
three-body forces in MeV. The number in bracket is binding energy
calculated in FY scheme in PW representation, Ref.
\cite{Kamada-NPA548}.}
\begin{tabular}{cc}
\hline
Potential  & Four-body Binding Energy  \\
\hline
MT-V            &   -31.3 [-31.36] \\
MT-V+MT3-I      & -38.8\\
MT-V+MT3-II     & -37.5\\
\hline
\end{tabular}
\end{table}

For four-body (three-body) binding energy calculations thirty
(forty) grid points for Jacobi momentum variables and twenty
(thirty two) grid points for angle variables have been used
respectively. As demonstrated in tables \ref{table1} and
\ref{table2}, the calculations of the three- and four-body binding
energies using only the MT-V 2BF yield the values $E=-7.74$ and
$-31.3 \, {\mathrm{MeV}}$, Ref. \cite{Hadizadeh-FBS40}.

In our calculations for 3BF we use a model of 3BF which is based
on multi-meson exchanges. We study two different types of 3BFs, a
purely attractive and a superposition of attractive and repulsive,
which are named MT3-I and MT3-II respectively Ref.
\cite{Liu-FBS33}. As shown in Ref. \cite{Liu-FBS33} The parameters
of these 3BFs are chosen so that the correction due to these 3BFs
to the three-body binding energy calculated with the modified
Malfliet-Tjon 2BF (MT2-II) is small, and they lead to binding
energies near to the triton binding energy.

The three- and four-body binding energies calculated in 3D
approach are given in tables \ref{table1} and \ref{table2}. Our
results for three-body binding energies with the addition of the
MT3-I and MT3-II 3BFs, while MT-V is used as 2BF, are $-8.92$ and
$-8.70$ [MeV] and while MT2-II is used as 2BF are $-8.87$ and
$-8.64$ respectively. Our results agree with corresponding values
presented in Ref. \cite{Liu-FBS33} and \cite{Elster-FBS27}. Our
results for four-body binding energies with the addition of the
MT3-I and MT3-II 3BFs, while MT-V is used as 2BF, are $-38.8$ and
$-37.5$ [MeV] respectively. Unfortunately we could not compare our
results for four-body binding energies with other calculations,
since to the best of our knowledge no comparable work with scalar
two-meson exchange 3BFs exists. So in order to test the accuracy
of our calculations we carried out two numerical tests which are
presented in next section.

According to our experience for four-body bound state calculations
with 2BF alone \cite{Hadizadeh-FBS40}, we expect that our results
with 3BF provide the same accuracy in comparison to other
calculations of the four-body binding energy based on PW
decomposition, while the numerical procedure are actually easier
to implement.

\subsection{Test of Calculations}
In this section we investigate the numerical stability of our
algorithm and our 3D representation of Yakubovsky components. We
specially investigate the stability of the eigenvalue of the
Yakubovsky kernel with respect to the number of grid points for
Jacobi momenta, polar and azimuthal angle variables. We also
investigate the quality of our representation of the Yakubovsky
components and consequently wave function by calculating the
expectation value of the Hamiltonian operator.

In table \ref{table3} we present the obtained eigenvalue results
for binding energies given in tables \ref{table1} and \ref{table2}
for different grids. We choose the number of grid points for
Jacobi momenta as $N_{jac}$, for spherical angles as $N_{sph}$ and
for polar angles as $N_{pol}$. As demonstrated in this table, the
calculation of the eigenvalues $\lambda$ convergence to the value
one for $N_{jac}=30$ and $N_{sph}=N_{pol}=20$. It should be clear
that the solution of coupled Yakubovsky three-dimensional integral
equations, with six independent variables for the amplitudes, is
much more time-consuming with respect to the solution of
three-dimensional Faddeev integral equation \cite{Liu-FBS33}, with
three variables for the amplitude.

\begin{table}[hbt]
\caption{Stability of the eigenvalue $\lambda$ of Yakubovsky kernel
with respect to the number of grid points in Jacobi momenta
$N_{jac}$, spherical angles $N_{sph}$ and polar angles $N_{pol}$.
$E_{MT-V}=-31.3$, $E_{MT-V+MT3-I}=-38.8$, $E_{MT-V+MT3-II}=-37.5$
MeV and $\lambda_{1}, \lambda_{2}$ and $\lambda_{3}$ are
corresponding eigenvalues.}
\begin{tabular} {ccccccc}
\hline
$N_{jac}$  & $N_{sph}=N_{pol}$ & $\lambda_{1}$  & $\lambda_{2}$ & $\lambda_{3}$ \\
\hline
20 & 20 & 0.987 & 0.988 & 1.010\\
26 & 20 & 0.995 & 0.996 & 1.004\\
30 & 12 & 0.997 & 0.997 & 1.003\\
30 & 16 & 0.999 & 0.999 & 1.001\\
30 & 20 & 1.000 & 1.000 & 1.000\\
\hline
 \end{tabular} \label{table3}
\end{table}

The solution of coupled Yakubovsky three-dimensional integral
equations in momentum space allows to estimate numerical errors
reliably. With the binding energy $E$ and the Yakubovsky
components $|\psi_{1}\rangle $ and $|\psi_{2}\rangle $ available,
we are able to calculate the total wave function $|\Psi\rangle $
from Eq.~(\ref{Eq.WF}) by considering the choice of coordinate
systems which are represented in Ref. \cite{Hadizadeh-FBS40}. So
in order to demonstrate the reliability of our calculations we can
evaluate the expectation value of the Hamiltonian operator $H$ and
compare this value to the previously calculated binding energy of
the eigenvalue equation, Eq.~(\ref{Eq.35}). Explicitly we evaluate
the following expression:
\begin{eqnarray}
\langle \Psi |H| \Psi \rangle &=&  \langle \Psi |H_0| \Psi \rangle
      +    \langle \Psi | V | \Psi \rangle
      +    \langle \Psi | W | \Psi \rangle
\nonumber \\* &=&  12 \,\langle \psi_{1} |H_0| \Psi \rangle
      + 6 \,\langle \psi_{2} |H_0| \Psi \rangle
   \nonumber \\*   && + \, 6 \,\langle \Psi | V_{12} | \Psi \rangle
   \nonumber \\*   && + \,  4 \,\langle \Psi | W_{123} | \Psi \rangle
 \label{Eq.37}
\end{eqnarray}
where $V$ represents the 2BFs $(\sum_{i<j} V_{ij})$ and $W$ the 3BFs
$(\sum_{i<j<k} W_{ijk})$. The expectation value of the kinetic
energy $\langle H_0 \rangle$ and the 2B potential energy $\langle
V_{12} \rangle$ have been evaluated in Ref. \cite{Hadizadeh-FBS40}.
The expectation value of the 3B potential energy, $\langle W_{123}
\rangle$, is given by
\begin{eqnarray}
\langle \Psi | W_{123} | \Psi \rangle &=& 3\langle \Psi |
W_{123}^{(3)} | \Psi \rangle \nonumber
  \\ &=& 3\times8 \pi^2 \int_{0}^{\infty} du_{1} \, u^{2}_{1}
\int_{-1}^{+1} dx_{1} \int_{0}^{2\pi} d\varphi_{1} \nonumber
  \\ &\times& \int_{0}^{\infty}
du_{2} \, u^{2}_{2} \int_{-1}^{+1} dx_{2} \int_{0}^{\infty} du_{3}
\, u^{2}_{3} \nonumber
  \\ &\times&  \Psi(u_{1}\,u_{2}\,u_{3}\,x_{1}\,x_{2}\,
\varphi_{1} )  W_{123}^{(3)} \,
  \Psi(u_{1}\,u_{2}\,u_{3}\,x_{1}\,x_{2}\,
\varphi_{1} ) \nonumber
  \\ \label{Eq.38}
\end{eqnarray}
Here the integrations need the evaluation of the matrix element
$_{3}\langle\vec{u}\,_{1}\,\vec{u}\,_{2}\,\vec{u}\,_{3}|W_{123}^{(3)}
|\Psi\rangle$ of Eq.~(\ref{Eq.20}). The expectation values of the
kinetic energy $\langle H_0\rangle$, the 2B interaction $\langle
V\rangle$, the 3B interaction $\langle W\rangle$ and the Hamiltonian
operator $\langle H \rangle$ for three- and four-body bound states
are given in tables \ref{table4} and \ref{table5} respectively. In
the same tables the corresponding binding energies calculated in 3D
scheme are also shown for comparison to the expectation values of
the Hamiltonian operator. One can see that the energy expectation
value and eigenvalues $E$ agree with high accuracy. All these
numbers are not meant to provide insight into the physics of three
and four interacting nucleons, but serve only as a demonstration
that this technique allows a very accurate and easy handling of
typical nuclear forces consisting of attractive and repulsive (short
range) parts. In addition, they will serve as benchmarks for future
studies.

\begin{table}[hbt]
\caption{Expectation values with respect to the three-body wave
functions for various potential combinations. We present the
expectation values of the kinetic energy $\langle H_{0}\rangle$, the
2B interaction $\langle V\rangle$ and the three-body interaction
$\langle W\rangle$. Additionally the expectation values of the
Hamiltonian operator $\langle H\rangle$ are compared to the binding
energy results from the Faddeev equations. All energies are given in
MeV.}
\begin{tabular} {cccccccccccc}
\hline
 Potential & $\langle H_0\rangle$ & $\langle V\rangle$ & $\langle W\rangle$ & $\langle H\rangle$ & $E$ &
\\   \hline
MT-V & 29.77 & -37.51 & - & -7.74 & -7.74 &
 \\
MT-V+MT3-I & 33.13 & -40.63 & -1.41 & -8.91 & -8.92 &
\\
MT-V+MT3-II & 32.38 & -40.02 & -1.07 & -8.71 & -8.70 &
\\
MT2-II & 28.64 & -36.33 & - & -7.69 & -7.69 &
 \\
MT2-II+MT3-I & 31.88 & -39.40 & -1.34 & -8.86 & -8.87 &
\\
MT2-II+MT3-II & 31.17 & -38.78 & -1.04 & -8.65 & -8.64 &
\\ \hline
\end{tabular} \label{table4}
\end{table}

\begin{table}[hbt]
\caption{The same as table \ref{table4}, but for four-body case.}
\begin{tabular} {cccccccccccc}
\hline
 Potential & $\langle H_0\rangle$ & $\langle V\rangle$ & $\langle W\rangle$ & $\langle H\rangle$ & $E$ &
\\  \hline
MT-V & 69.7 & -101.0 & - & -31.3 & -31.3 &
 \\
MT-V+MT3-I & 78.8 & -110.1 & -7.5 & -38.8 & -38.8 &
\\
MT-V+MT3-II & 76.1 & -107.6 & -6.0 & -37.5 & -37.5 &
\\ \hline
\end{tabular} \label{table5}
\end{table}

3BF effects have a stronger impact on four-body bound state than for
three-body bound state as can be seen for instance by comparing
expectation values of the potential energies for the two systems. We
find in case of MT-V $\langle V\rangle=-37.51$ ($-101.0$) MeV for
three (four)-body bound state without 3BF and $\langle
V\rangle=-40.63, -40.02$ ($-110.1, -107.6 $) MeV with MT3-I and
MT3-II 3BFs correspondingly. In the latter case the expectation
values for the 3BFs are $\langle W\rangle=-1.41, -1.07$ ($-7.5 ,
-6.0$) MeV for three (four)-body bound state. Already the trivial
fact that there are four triplets in four-body bound state makes it
clear that one has to expect 3BF effects to be more pronounced in
the four-body bound state than in the three-body bound state.

\section{Summary and Outlook}\label{section: summary}

Instead of solving the coupled Faddeev-Yakubovsky equations in a
partial wave basis, we introduce an alternative approach for
four-body bound state calculations which implement directly
momentum vector variables. We formulated the coupled Yakubovsky
equations for identical spinless particles, interacting by two-
and three-body forces, as function of vector Jacobi momenta,
specifically the magnitudes of the momenta and the angles between
them. We expect that coupled three-dimensional Yakubovsky
equations for a bound state can be handled in a straightforward
and numerically reliable fashion. In comparison to an angular
momentum decomposition which is commonly used
\cite{Kamada-NPA548}-\cite{Epelbaum-PRC70}, this direct approach
has great advantages. In our Three-Dimensional case there is only
two coupled three-dimensional integral equations to be solved,
whereas in the partial wave case one has two coupled sets of a
finite number of coupled equations with kernels containing
relatively complicated geometrical expressions. The comparison of
3D and PW formalisms shows that our 3D formalism avoids the very
involved angular momentum algebra occurring for the permutations
and transformations and it is more efficient especially for the
three-body forces.

The three dimensional Yakubovsky integral equations was
successfully solved using Malfliet-Tjon type 2BF alone, and its
numerical feasibility and accuracy established
\cite{Hadizadeh-FBS40}. Here we present results including the
scalar two-meson exchange three-body force and study its effects
on the energy eigenvalue and the four-body wave function. The
stability of our algorithm and our Three-Dimensional
representation of Yakubovsky components have been achieved with
the calculation of the eigenvalue of Yakubovsky kernel, where
different number of grid pints for Jacobi momenta and angle
variables have been used. Also we have calculated the expectation
value of the Hamiltonian operator. This test of calculation
represents good agreement between the obtained eigenvalue energy
and expectation value of the Hamiltonian operator.

This is very promising and nourishes our hope that calculations
with realistic two and three-nucleon forces, namely considering
spin and isospin degrees of freedom, will most likely be more
easily implemented than the traditional partial wave based method.

To this aim the first step for realistic calculations of
three-nucleon bound state in a realistic Three-Dimensional
approach has already been taken by calculation of Triton binding
energy with Bonn-B potential \cite{Bayegan-EFB20,Bayegan-PRC} and
formulation of four-nucleon bound state is currently underway and
it will be reported elsewhere \cite{Hadizadeh-in_preperation}.
They will be the first steps for realistic calculations of three-
and four-nucleon bound states in a Three-Dimensional scheme.

It should be mentioned that the input to such calculations is the
NN $t$-matrix which is calculated in an approach based on a
helicity representation and depends on the magnitudes of the
initial and final momenta and the angle between them
\cite{Fachruddin-PRC62}. Consequently the calculation of NN
$t$-matrix in helicity representation needs the NN potentials in
an operator form which can be incorporated in 3D formalism. As
indicated in sec. $3.2$ of Ref. \cite{Fachruddin-PhD} (or sec. III
of Ref. \cite{Fachruddin-PRC62}) the general structure of the NN
potential operator which fits well to the helicity representation
is given, and on this representation both Bonn-B and AV18 NN
potentials are given in operator form, see appendixes C and D (or
sec. IV of Ref. \cite{Fachruddin-PRC62}).

\section*{Acknowledgments}
One of authors (M. R. H.) would like to thank H. Kamada for
fruitful discussions about three-body forces during APFB05
conference. This work was supported by the research council of the
University of Tehran.



\begin{thebibliography}{0}

\bibitem{Primakoff-PR55} H. Primakoff, T. Holstein, {\it Phys. Rev.} {\bf 55}, 1281
(1939).

\bibitem{Robilotta-FBS35} M. R. Robilotta, {\it Few Body Syst., Suppl.} {\bf 2}, 35
(1987).

\bibitem{Fujita-PTP17} J. Fujita et al., {\it Progr. of Theor. Phys.} {\bf 17}, 360
(1957).

\bibitem{Meissner-NPA684} U. G. Mei{\ss}ner, E. Epelbaum and W. Gl\"{o}ckle, {\it Nucl. Phys.} {\bf A 684}, 371
(2001).

\bibitem{Epelbaum-PRC66} E. Epelbaum et al., {\it Phys. Rev.} {\bf C 66}, 064001
(2002).

\bibitem{Epelbaum-NPA747} E. Epelbaum, U. G. Mei{\ss}ner and W. Gl\"{o}ckle, {\it Nucl. Phys.} {\bf A 747},
362 (2005).

\bibitem{Epelbaum-PRC71} E. Epelbaum, U. G. Mei{\ss}ner and  J. E. Palomar, {\it Phys. Rev.} {\bf C 71},
024001 (2005).

\bibitem{Epelbaum-PLB639} E. Epelbaum, {\it Phys. Lett.} {\bf B 639},
456 (2006).

\bibitem{Epelbaum-PPNP57} E. Epelbaum, {\it Prog. Part. Nucl. Phys.} {\bf 57},
654 (2006).

\bibitem{Bedaque-NPA676}  P. F. Bedaque, H.-W. Hammer and U. van Kolck, {\it Nucl. Phys.} {\bf A 676},
357 (2000).

\bibitem{Hiyama-PRL85} E. Hiyama et al., {\it Phys. Rev. Lett.} {\bf 85}, 270 (2000).

\bibitem{Usukura-PRB59} J. Usukura, K. Varga and Y. Suzuki, {\it Phys. Rev.} {\bf B 59}, 5652 (1999).

\bibitem{Viviani-PRC71} M. Viviani, A. Kievsky and S. Rosati, {\it Phys. Rev.} {\bf C 71}, 024006 (2005).

\bibitem{Viringa-PRC62} R. B. Viringa et al., {\it Phys. Rev.} {\bf C 62}, 014001 (2000).

\bibitem{Navratil-PRC62} P. Navr\'{a}til, J. P. Vary and B. R. Barret, {\it Phys. Rev. } {\bf C 62}, 054311 (2000).

\bibitem{Barnea-PRC67} N. Barnea, W. Leidemann and G. Orlandini, {\it Phys. Rev.} {\bf C 67}, 054003 (2003).

\bibitem{Schellingerhout-PRC46} N. W. Schellingerhout, J. J. Schut, and L. P. Kok, {\it Phys. Rev.}
{\bf C 46}, 1192 (1992).

\bibitem{Lazauskas-PRC7} R. Lazauskas and J. Carbonell, {\it Phys. Rev.}
{\bf C 70}, 044002  (2004).

\bibitem{Kamada-NPA548} H. Kamada and W. Gl\"{o}ckle, {\it Nucl. Phys.} {\bf A 548},
205 (1992).

\bibitem{Glockle-NPA560} W. Gl\"{o}ckle and H. Kamada, {\it Nucl. Phys.} {\bf A 560},
541 (1993).

\bibitem{Nogga-PRL85} A. Nogga, H. Kamada and W. Gl\"{o}ckle, {\it Phys. Rev. Lett.} {\bf 85},
944 (2000).

\bibitem{Nogga-PHD} A. Nogga, Ph.D. thesis, Ruhr-Universit\"{a}t, Buchum (2001).

\bibitem{Kamada-PRC64} H. Kamada et al., {\it Phys. Rev.} {\bf C 64},
044001 (2001).

\bibitem{Nogga-PRC65} A. Nogga, H. Kamada, W. Gl\"{o}ckle and B. R.
Barrett, {\it Phys. Rev.} {\bf C 65}, 054003 (2002).

\bibitem{Epelbaum-PRL86} E. Epelbaum et al., {\it Phys. Rev. Lett.} {\bf 86}, 4787
(2001).

\bibitem{Epelbaum-EPJA15} E. Epelbaum et al., {\it Eur. Phys. J.} {\bf A 15 }, 543
(2002).

\bibitem{Epelbaum-PRC70} E. Epelbaum et al., {\it Phys. Rev.} {\bf C 70}, 024003
(2004).

\bibitem{Elster-FBS24} Ch. Elster, J. H. Thomas, W. Gl\"{o}ckle, {\it Few Body Syst.} {\bf 24}, 55 (1998).

\bibitem{Elster-FBS27} Ch. Elster, W. Schadow, A. Nogga, W. Gl\"{o}ckle, {\it Few Body Syst.} {\bf 27}, 83 (1999).

\bibitem{Schadow-FBS28} W. Schadow, Ch. Elster, W. Gl\"{o}ckle, {\it Few Body Syst.} {\bf 28}, 15 (2000).

\bibitem{Fachruddin-PRC62} I. Fachruddin, Ch. Elster, W. Gl\"{o}ckle, {\it Phys. Rev.} {\bf C 62}, 044002 (2000).

\bibitem{Fachruddin-PRC63} I. Fachruddin, Ch. Elster, W. Gl\"{o}ckle, {\it Phys. Rev.} {\bf C 63}, 054003 (2001).

\bibitem{Liu-FBS33} H. Liu, Ch. Elster, W. Gl\"{o}ckle, {\it Few Body Syst.} {\bf 33},
241 (2003).

\bibitem{Fachruddin-MPLA18} I. Fachruddin, Ch. Elster, W. Gl\"{o}ckle, {\it Mod. Phys. Lett.} {\bf A 18}, 452 (2003).

\bibitem{Fachruddin-PRC68} I. Fachruddin, Ch. Elster, W. Gl\"{o}ckle, {\it Phys. Rev.} {\bf C 68}, 054003 (2003).

\bibitem{Fachruddin-PRC69} I. Fachruddin, W. Gl\"{o}ckle, Ch. Elster, A. Nogga, {\it Phys. Rev.} {\bf C 69}, 064002 (2004).

\bibitem{Liu-PRC72} H. Liu, Ch. Elster, W. Gl\"{o}ckle, {\it Phys. Rev.} {\bf C 72}, 054003 (2005).

\bibitem{Lin-PRC76} T. Lin, Ch. Elster, W. N. Polyzou, and W. Gl\"{o}ckle, {\it Phys. Rev.} {\bf C 76}, 014010  (2007).

\bibitem{Hadizadeh-WS} M. R. Hadizadeh and S. Bayegan, {\it Proceedings of the 3rd Asia-Pacific
Conference, Nakhon Ratchasima, Thailand, July 2005} (World
Scientific, Singapore, 2007, p. 16).

\bibitem{Hadizadeh-FBS40} M. R. Hadizadeh and S. Bayegan, {\it Few Body Syst.} {\bf 40},
171 (2007).

\bibitem{Huber-FBS22} D. H\"{u}ber, H. Witala, A. Nogga, W. Gl\"{o}ckle and H. Kamada, {\it Few Body Syst.} {\bf 22}, 107 (1997).

\bibitem{Stadler-PRC44} A. Stadler, W. Gl\"{o}ckle and P. U. Sauer, {\it Phys. Rev.} {\bf C 44}, 2319 (1991).

\bibitem{Malfliet-NPA127} R. A. Malfliet and J. A. Tjon, {\it Nucl. Phys.} {\bf A 127},
161 (1969).

\bibitem{Bayegan-EFB20} S. Bayegan, M. R. Hadizadeh and M. Harzchi, {\it to appear in Few Body Syst. as
proceedings of the 20th European Conference on Few-Body Problems
in Physics, Pisa, Italy, Sep. 2007}. {\it arXiv:0711.4036}.

\bibitem{Bayegan-PRC} S. Bayegan, M. R. Hadizadeh and M. Harzchi, {\it submitted to Phys. Rev.} {\bf
C}. {\it arXiv:0711.4026 }.

\bibitem{Hadizadeh-in_preperation} M. R. Hadizadeh and S. Bayegan, {\it in preparation}.

\bibitem{Fachruddin-PhD} I. Fachruddin, PhD. thesis, Ruhr-Universit\"{a}t Bochum, 2002.

\end{thebibliography}
\end{document}